# Achievements and Lessons Learned from Successful Small Satellite Missions for Space Weather-Oriented Research

**Harlan E. Spence[1], Amir Caspi[2], Hasan Bahcivan[3], Jesus Nieves-Chinchilla[4], Geoff Crowley[5], James Cutler[6], Chad Fish[5], David Jackson[7], Therese Moretto Jorgensen[8], David Klumpar[9], Xinlin Li[10], James P. Mason[10,11], Nick Paschalidis[12], John Sample[9], Sonya Smith[1], Charles M. Swenson[13], and Thomas N. Woods[10]**

[1]Institute for the Study of Earth, Oceans, and Space, University of New Hampshire, Durham, NH

[2]Southwest Research Institute, Boulder, CO

[3]SRI International, Menlo Park, CA

[4]Technical University of Madrid, Spain

[5]Atmospheric & Space Technology Research Associates (ASTRA) LLC, Boulder, CO

[6]University of Michigan, Ann Arbor, MI

[7]UK Met Office, United Kingdom

[8]NASA Ames Research Center, Mountain View, CA

[9]Montana State University, Bozeman, MT

[10]Laboratory for Atmospheric and Space Physics, University of Colorado, Boulder, CO

[11]Applied Physics Laboratory, The Johns Hopkins University, Laurel, MD

[12]NASA Goddard Space Flight Center, Greenbelt, MD

[13]Utah State University

Corresponding author: Harlan E. Spence ([Harlan.Spence@unh.edu](Harlan.Spence@unh.edu))

**Key Points:**

- Many NSF/NASA-funded CubeSat missions have contributed significantly to space weather research and applications
- Low-cost missions benefit from a rapid "fly-learn-modify-refly" cycle
- CubeSat science productivity is comparably high to larger missions if normalized by cost or by weighted impact of refereed publications

**Abstract**

When the first CubeSats were launched nearly two decades ago, few people believed that the miniature satellites would likely prove to be a useful scientific tool. Skeptics abounded. However, the last decade has seen the highly successful implementation of space missions that make creative and innovative use of fast-advancing CubeSat and small satellite technology to carry out important science experiments and missions. Several projects now have used CubeSats to obtain first-of-their-kind observations and findings that have formed the basis for high-profile engineering and science publications, thereby establishing without doubt the scientific value and broad utility of CubeSats. In this paper, we describe recent achievements and lessons learned from a representative selection of successful CubeSat missions with a space weather focus. We conclude that these missions were successful in part because their limited resources promoted not only mission focus but also appropriate risk-taking for comparatively high science return. Quantitative analysis of refereed publications from these CubeSat missions and several larger missions reveals that mission outcome metrics compare favorably when publication number is normalized by mission cost or if expressed as a weighted net scientific impact of all mission publications.

**Plain Language Summary**

Space missions using very small satellites and low resources have demonstrated they can accomplish high quality science, overcoming initial low expectations of many inside the space science community. We focus on one class of small satellites known as "CubeSats". CubeSats comprise a small number of modular cubes, each the size of a typical tissue box and weighing approximately one kilogram (like a pineapple). We discuss five CubeSat missions that operated during the last ten years, each having total mission mass of three kilograms and total mission costs of slightly more than one million US dollars. These missions had focused goals targeting different aspects of space weather. For each mission, we summarize its scientific achievements and lessons learned, many of them common lessons. Larger missions have flown during this same time with overall mass ranging from hundreds to thousands of kilograms and mission costs many hundreds of thousands to over one billion US dollars. We compare the relative science value of these smallest and larger missions through the publications they produce in professional journals. Though CubeSat missions yield far fewer total publications compared to larger missions, the cost per publication is lower while still producing comparably high scientific impact.

## 1 Introduction

Increasingly, private industry as well as federal agencies in the US, including the Department of Defense (DoD), the National Aeronautics and Space Administration (NASA), and the National Science Foundation (NSF), are taking a serious look at CubeSats as a viable, low-cost option for space missions to help fulfill their respective needs. International agencies worldwide are also considering expanding their scientific goals with CubeSats and other small satellites (collectively, "smallsats"; in this paper, CubeSat and smallsat may be used interchangeably). We note that standardized containerization of CubeSats has proven to be one important element of the success of this platform. Ongoing and future CubeSat mission outcomes aim not only at advancing scientific research, but also at accomplishing other programmatic goals, such as surveillance and environmental monitoring. In addition, CubeSat projects provide essential opportunities to train the next generation of experimental scientists and engineers. CubeSat missions typically have limited scope, which generally enables a relatively rapid development and short ($\lesssim 1$ year) operational period. They therefore allow students and early-career professionals, through hands-on work on real-world, end-to-end projects, to develop the necessary skills and experience needed to succeed in Science, Technology, Engineering, and Mathematics (STEM) careers. CubeSat projects are also an effective tool to broaden the participation amongst underrepresented groups in STEM research, education, and workforce development. The projects stimulate widespread excitement and involve a uniquely diverse set of skills and interest. Therefore, they appeal to a broader range of participants than more traditional science and engineering projects.

In this paper, we review five recent CubeSat missions which produced significant outcomes and scientific results despite their limited scope and cost. Though these missions were led at institutions with rich prior experience in the development of spaceflight hardware, and though some PIs had prior experience in developing instrumentation for spaceflight, these were the first missions to be led by every PI, many of whom were early career; project teams typically included a diverse set of students and partner institutions comparatively newer to space missions. The NSF CubeSat program, managed by the Atmosphere and Geospace Section (AGS) of the Geosciences Directorate, supported all but the last of these missions; the final smallsat mission comes from a NASA program that began within the past decade, after the NSF program. While there are many other missions that could have been chosen, we chose ours for two primary reasons (see also Caspi et al., 2021): five missions allows for a representative and succinct sampling across space weather disciplines; and we focused on those missions presented at the 1st International Workshop on SmallSats for Space Weather Research and Forecasting, held in Washington, DC on 1–4 August 2017.

A key point of this paper is to highlight that CubeSats can indeed produce significant, quality science. We acknowledge that not all CubeSat missions will do so, but that is the nature of exploratory/developmental research at very low cost, which is what CubeSats are doing. A common element unifies these successful CubeSat missions regardless of the funding agency: goals meant to lead to better understanding of space weather or demonstrate potential application to space weather operational needs. We develop a set of metrics to quantify CubeSat mission success in terms of the refereed publications they produce. Finally, we compare the metrics of the selected successful CubeSat missions with those from a representative sample of larger successful NASA missions to demonstrate the scientific potential of both platform scales.

**2 Background and context**

The report entitled "Achieving Science with CubeSats: Thinking Inside the Box" (National Academies, 2016) provides an excellent overview on the status and evolution of CubeSats at that time. A committee of the National Academy of Sciences, Engineering, and Mathematics (NASEM) Space Studies Board (SSB) wrote this report at a time when CubeSat mission developments were still few though increasing at a rapid pace. Through workshops and other forms of data collection, the committee drew their conclusions from a broad swath of the CubeSat community, including, but not limited to: agencies that support mission development (e.g., NSF, NASA, DoD); organizations funded to develop and implement missions (e.g., universities, non-profits, government, private); entities who provide relevant regulatory oversight (e.g., National Oceanographic and Atmospheric Administration, Federal Communication Commission, Federal Aviation Administration, National Telecommunications and Information Administration); parts of the government that develop relevant policy (e.g., Office of Science and Technology Policy); and providers of launches and launch services. That contemporary report provides an excellent history of CubeSats and so we direct interested readers to that document for the remarkable story of the genesis and subsequent explosive growth of CubeSat missions. The report also provides "recommendations for near-term actions as well as on strategies for enhancing the scientific usefulness of CubeSats without overly restraining the spirit of innovation that characterizes the broad community of CubeSat users." We believe that the recommendations with broad community-consensus developed for their study remain valid to date and so again we refer the reader to their report for those important recommendations.

As noted above, the NASEM/SSB activities summarized an extremely broad array of topics required for a full assessment. Even though their report includes over 100 pages in total, given its breadth, the report necessarily could not also provide great depth, particularly in the sections summarizing missions and, even more so, in those sections summarizing specifically space science missions that focused on space weather. The report mentions space weather CubeSat mission descriptions, outcomes, and lessons learned only briefly. Furthermore, more than six years of space weather CubeSat activities have transpired since formulation of the NASEM/SSB report.

Accordingly, in this paper, we provide an updated summary of space weather-related CubeSat missions, focusing on five successful NSF- and NASA-funded missions. We consider two aspects for each mission: (1) a summary of scientific achievements, and (2) lessons learned. Section 3 provides detailed, updated summaries of these two aspects for each of the five missions, as relevant, listed in order of their launch dates. Those missions (often a series of related missions) comprise by order of launch date, then alphabetically if the same launch date: DICE, RAX-2 (and RAX-1), CSSWE, FIREBIRD-II (and FIREBIRD-I), and MinXSS-1 (and MinXSS-2). In Section 4, we discuss common themes and a quantitative analysis of scientific productivity of these missions compared to larger missions. In Section 5 we provide concluding remarks.

## 3 Updates of Five Space Weather-Themed CubeSat/SmallSat Missions

### *3.1 Dynamic Ionosphere CubeSat Experiment (DICE)*

DICE Overview

The Dynamic Ionosphere CubeSat Experiment (DICE) mission represents the first constellation of CubeSats executed specifically for scientific purposes. DICE was selected in October 2009 as part of NSF's inaugural "CubeSat-based Science Mission for Space Weather and Atmospheric Research" program. Like many of the early successful CubeSat missions in the NSF program, DICE was a collaborative effort, involving consortium members drawn from industry, government, and university partners. The DICE PI, Dr. Geoffrey Crowley of Atmospheric and Space Technology Research Associates (ASTRA LLC), and Deputy PI, Dr. Charles Swenson of Utah State University Space Dynamics Laboratory, developed the dual CubeSats (DICE-1 and DICE-2, nicknamed Farkle and Yahtzee, respectively) with other university partners at Embry-Riddle Aeronautical University and Clemson University, and industry partners at L-3 Communications, TiNi Aerospace, Clyde Space, Orbital ATK, and Pumpkin, Inc. as well as with NASA Goddard Space Flight Center. DICE mission overviews are provided in Crowley et al (2010; 2011) and Fish et al. (2012); the mission description is detailed in Fish et al. (2014).

DICE Scientific Achievements

The science of the DICE mission focused on a phenomenon known as Storm Enhanced Density (SED). SED is a process that produces large density gradients in the upper region of Earth's ionized upper atmosphere, called the ionosphere, which in turn leads to undesirable space weather conditions. Many critical systems rely on reliable radio frequency (RF) transmissions using the conducting ionosphere (e.g., communications, surveillance, and navigation). One form of space weather is thus the natural variability of the ionosphere. That variability can have dramatic effects on the operation of these systems. Prior to the DICE mission, understanding of the SED phenomenon was largely limited to remote sensing techniques.

The DICE mission had three scientific objectives:

1. Investigate the physical processes responsible for formation of the mid-latitude ionospheric SED bulge in the noon-to-post-noon sector during magnetic storms.

2. Investigate the physical processes responsible for the formation of the SED plume at the base of the SED bulge and the transport of the high-density SED plume across the magnetic pole.

3. Investigate the relationship between penetration of electric fields and the formation and evolution of SED.

Student teams (a total of 60 students overall) at each university in the consortium, in concert with senior members at all partners, were involved in the full life cycle of DICE, spanning the design, development, testing, and operation of the spacecraft as well as in the processing and analysis of the data. Starting with launch in October 2011, the DICE science team achieved full mission success in its two years of successful operations (before the RF transmit license from the International Telecommunications Union expired). Over that time, DICE provided the first CubeSat observations of the SED process in the ionosphere (Crowley et al., 2015). DICE measurements clarified how plasma density enhancements were transported into the polar cap from lower latitudes as part of the high-latitude convection pattern.

Other scientific successes relate as much to technical achievements (such as attitude determination described in Jandak and Fullmer, 2011; Ryan et al., 2011; and Neilsen et al., 2014) as to the aforementioned science objectives. For instance, DICE provided the first demonstration of how a body-mounted (i.e., boomless) magnetometer on a CubeSat could be used to infer ionospheric field-aligned currents (FACs). This activity led to a re-analysis of magnetic field aligned current (FAC) measurements from the NSF AMPERE mission when differences were found between the Active Magnetosphere and Planetary Electrodynamics Response Experiment (AMPERE) and DICE data; that discrepancy was subsequently resolved by comparison with Defense Meteorological Satellite Program (DMSP) observations (Delores Knipp, private communication). In addition, the collaboration between the DICE team and L3 Communications led to the commercial "Cadet" UHF nanosat radio. The Cadet radio provided a high-speed communications link with unprecedented data rates (3 Mbit/s downlink) for this class of spacecraft. The high data rates enabled much larger amounts of data to be recovered from the DICE CubeSats than had previously been possible from the typical 9600 kbit/s UHF downlink speeds of prior missions.

DICE Lessons Learned

As the Cadet radio development noted above underscores, CubeSats may be small, but their data volume need not be. Small satellites can still generate large amounts of data which require the same amount of careful analysis and quality control demanded on large missions. However, because CubeSats are resource constrained, often only a small fraction of data can be recovered. Given that CubeSats can generate a firehose of data but can generally only transmit through a soda straw, it is imperative to consider that upfront in mission design. We note that recent developments in S-band and higher frequency radios (discussed in the MinXSS section) provide opportunities for greater data recovery (see also Caspi et al., 2021). Both funding and time need to be included in the planning process to provide the level of resources needed for data analysis from CubeSat missions.

The second lesson learned from the DICE experience relates to deployables. Deployable items such as antennas and booms can add considerably to mission risk. If a mission can avoid a deployable, that is certainly desirable, though many current missions have used deployables successfully; repeated use of successful heritage deployable designs does mitigate risk. The body-mounted magnetometer provided compelling evidence that boomless magnetometer applications work adequately, at least for certain CubeSat applications.

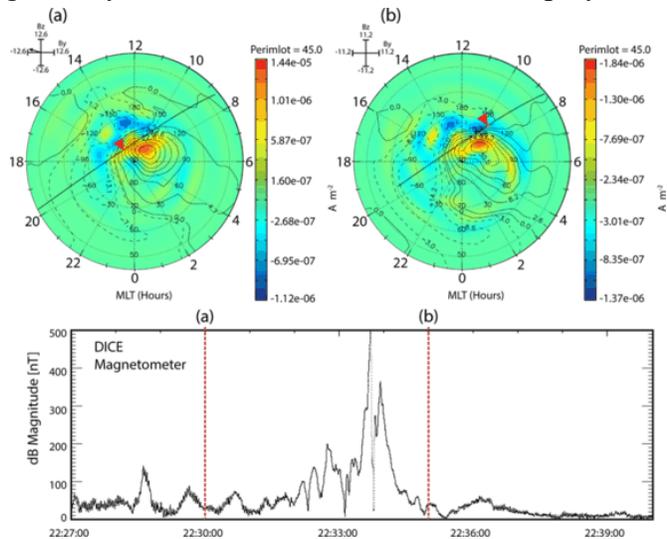

Figure 1, reproduced from Figure 52 of Fish et al. (2014), illustrates how well the body-mounted DICE magnetometer intensity measurements agree (lower panel) along the spacecraft trajectory in time with a data-driven assimilative model (upper panels) of dynamic electrical currents flowing into and out of the auroral regions and the disturbance magnetic fields they produce along the spacecraft orbit path.

*Figure 1. Comparison of AMIE model FAC's versus the Farkle SciMag dB magnitudes for the geomagnetic disturbance observed on May 22, 2012. (From Fish et al., 2014)*

### *3.2 Radio Aurora eXplorer-2 (RAX-2)*

RAX-2 Overview

The Radio Aurora eXplorer (RAX) mission was the first CubeSat mission launched under the NSF CubeSat program (Moretto, 2008). The two satellites of the mission (RAX-1 and RAX-2) studied high latitude space weather phenomena. RAX was a collaborative effort between SRI

International and the Michigan eXploration Laboratory (MXL) at the University of Michigan; Dr. Hasan Bahcivan of SRI developed the science payload while Professor James Cutler at MXL led a team to develop, build, and operate the satellite systems.   RAX-1 launched from Kodiak, AK on 19 November 2010 in collaboration with the Space Test Program operated by the US Department of Defense.   RAX-2 launched less than a year later on 28 October 2011 onboard a Delta-II as part of the NASA ELaNa-3 mission.   RAX successfully laid the foundation for the NSF CubeSat program and the follow-on space weather missions.

RAX-2 Science Achievements

The RAX mission studied how the aurora contributes to heating of the ionosphere (Cutler and Bahcivan, 2013].   This heat flow in the plasma of the ionosphere is an important process in space weather, which, if understood, will help us better understand how space weather impacts the ionosphere and the resulting communication challenges between Earth and space.   The specific RAX science objective was to study an important class of ionospheric disturbances, so called magnetic field-aligned irregularities (FAI).   These small, sub-meter size irregularities in the ionospheric plasma are known to disrupt communication signals.   Considerable effort has been made in the last two decades to model this heating process analytically and numerically (Bahcivan et al., 2009). However, although the models reproduce enhanced plasma temperatures well, a basic assumption underlying the theoretical models has never been verified experimentally, namely the degree to which observed FAIs are aligned with the magnetic field. Determination of this magnetic alignment sensitivity is critical not only for quantifying local plasma heating, but also for quantifying total heating rates in the ionosphere.

The RAX satellites were developed to measure the magnetic field alignment of FAIs (Bahcivan and Cutler, 2012).   Past remote sensing experiments were unable to measure this alignment due to the geometry of the radar sensing systems; the high latitude FAIs are aligned with the magnetic field, which is nearly vertical at high latitudes.   The radar transmissions reflect off into space rather than back to the ground-based radar transmitter.   RAX was thus designed to be a "bistatic" radar system by which we mean that the RAX satellites act as radar receivers, receiving signals from a network of ground-based, high-powered, northern radar transmitters.   The primary transmitter was the ground-based Poker Flat Incoherent Scatter Radar (PFISR) which illuminated the FAI with radio waves during overhead flights of the RAX satellites.

RAX enabled, for the first time, direct measurement of the magnetic alignment of the FAI (Bahcivan et al., 2012). Figure 2, a reproduction of Bahcivan et al's Figure 3, shows an example of the range-time-intensity image measured by the experiment and used in their analysis; the black curve shows the expected location of echoes originating from the altitude of 100 km compared to the measured signal (bright red patch between 230 to 240 seconds with a high SNR return. Based on several compelling measurements of this sort, the RAX science team discovered that the magnetic alignment sensitivity of the small FAIs is far higher than previously believed, contrary to what has been assumed in most models. This finding suggests that small scale waves are too electrically weak to contribute significantly to E-region electron heating. RAX results thus cast doubt on the decades-old theories about how plasma instabilities contribute to ionospheric heating. Rather, Bahcivan et al. (2014) conclude from these RAX results that the dynamics of decameter or longer wavelength FAIs significantly contribute to anomalous electron heating in the auroral region of Earth's ionosphere. These findings enable the creation of new models that better predict these ionospheric heating events and the conditions that spawn their creation.

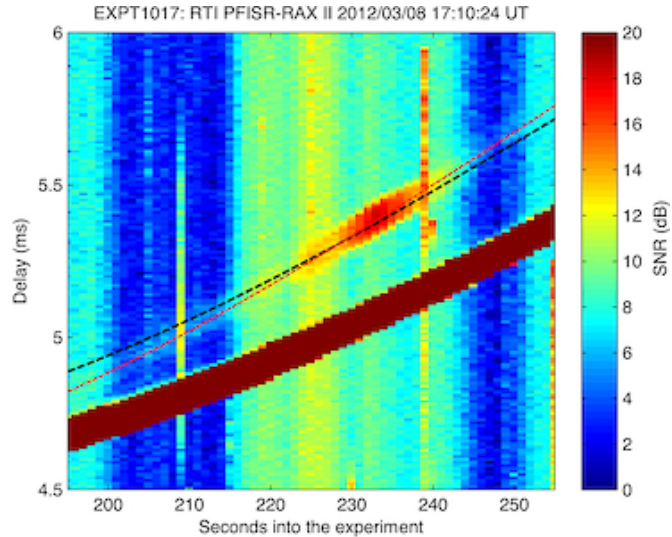

*Figure 2. Range-time-intensity plot for the duration of E region echoes observed by RAX. The black line marks the arrival time of echoes from the altitude of 100 km. The red line is a visual fit to the trace of the echo peak. (From Bahcivan et al., 2012)*

RAX-2 Lessons Learned

The first lesson learned is an important scientific one: until one makes a definitive measurement to test a theory, one should be skeptical of it even if it is decades-old conventional wisdom. RAX-2's unique measurement strategy enabled a definitive test of a key assumption of the prevailing theory and models. These groundbreaking measurements facilitated the necessary insight into the electrodynamics of ionospheric heating due to plasma waves and have cast into doubt a long-standing belief.

Second, RAX was the first CubeSat mission to prove that science can come from this small, standardized satellite form factor. Of all the CubeSat missions discussed in this paper, RAX-1 launched first of the group. RAX-1 underscores the rapid design/flight cycle of CubeSats; the RAX team only had one year to develop their first satellite. Furthermore, RAX-2 amply highlights the "fly-learn-modify-refly" cycle. Ultimately, RAX-2 performed a novel, focused science study that provided data to improve our understanding of ionospheric heating and the resulting instabilities that impact space communication.

An engineering lesson learned is that small teams adapting Agile practices can quickly overcome design flaws and challenges to produce functional systems (Springmann et al., 2012, 2014; Spangelo et al., 2013; Springmann and Cutler, 2014). Two RAX satellites were built due to a failure in RAX-1, whose mission ended after approximately two months of operation due to a gradual degradation of the solar panels that ultimately resulted in a loss of power. The MXL team, still present at Michigan after the RAX-1 launch, was able to quickly iterate on a design fix and launch a second RAX-2 within a year at about 10% cost of the original RAX-1 mission. CubeSats, small and standardized in size, enabled easy launch of the second system.

Operationally, the team learned that existing, low-cost resources can be used to improve data downlink (Spangelo et al., 2015). High speed, low-cost radios did not exist for CubeSats during RAX development. Instead, a low-cost, low-rate radio transmitting at 9600 bps was used in conjunction with an ad hoc, federated ground station network. Amateur operators around the world successfully relayed 4-10x more data than the primary station at MXL. Longer contact times were used instead of unavailable higher rates. This opened the trade space for CubeSats to leverage a variety of heterogenous communication systems.

### *3.3 Colorado Student Space Weather Experiment (CSSWE)*

CSSWE Overview

The Colorado Student Space Weather Experiment (CSSWE) was an NSF-funded 3U CubeSat. Professors Xinlin Li and Scott Palo at the University of Colorado Boulder served as the CSSWE PI and co-PI. The CSSWE team of students managed, designed (Gerhardt and Palo, 2010; Schiller et al., 2014), built, tested (Blum et al., 2012; Gerhardt and Palo, 2016), operated, and analyzed data (Li et al., 2012) from the CSSWE mission (Palo et al., 2010; Li et al., 2011; Li et al., 2013a). CSSWE was delivered in January 2012 and launched on 13 September 2012 out of Vandenberg Air Force Base as part of the NASA ELaNa VI launch. Arguably, the CSSWE mission is the most scientifically successful yet, of the NSF CubeSats, or any CubeSat program (see productivity metrics developed in Table 1 below). The CSSWE mission goals were threefold:

1. Develop a student-designed CubeSat system for space weather investigation

2. Understand the relationships between solar energetic protons (SEPs), flares, and coronal mass ejections (CMEs)

3. Characterize the variations of the Earth's radiation belt electrons

In our technological, space-based society, spaceborne electronic systems are vulnerable to the hostile space environments in which they operate. The Earth's inner magnetosphere contains a region known as the Van Allen radiation belts which is a particularly hostile environment. It is filled with so-called relativistic electrons (those with energies from hundreds of keV to multiple MeV). Electrons with these energies can easily penetrate shielding of either a space suit or electronic parts, and lead to ionizing radiation or internal charging; both pose space weather hazards. Despite the recent completion of NASA's Van Allen Probes (Mauk et al., 2013), a large mission dedicated to studying the belts, there remain unanswered questions about the process by which electrons enter and exit the radiation belts. Furthermore, coronal mass ejections (CMEs) and some solar flares produce high-energy solar energetic protons (SEPs), which are harmful to astronauts and electronics alike. By sensing the directional flux and energy of both relativistic electrons and protons, a connection may be drawn between solar events (flares and CMEs), radiation belt evolution, and SEPs. Understanding the coupled dynamics of these events is crucial to determining the effect of solar activity on satellite systems and developing strategies for predicting and mitigating the impacts.

With this scientific focus in mind, CSSWE was a strategically conceived single-instrument mission. It flew the Relativistic Electron and Proton Telescope integrated little experiment (REPTile; Schiller et al., 2010) to provide directional differential flux measurements of high-energy electrons and protons near the atmosphere, complementary to the REPT instrument (Baker et al., 2012) of the Radiation Belt Storm Probes – Energetic particle, Composition and Thermal plasma suite (Spence et al., 2013) of the Van Allen probes mission. CSSWE's high inclination orbit compared to Van Allen Probes equatorial orbit provided the critical opportunity to connect the radiation belts between low-Earth orbit (CSSWE) and medium-Earth orbit (Van Allen Probes).

CSSWE Science Achievements

CSSWE's measurements have helped us understand better the loss of relativistic electrons from the radiation belts to the upper atmosphere. Though a tiny fraction of the cost of Solar Anomalous and Magnetospheric Particle Explorer (SAMPEX), the first of NASA's Small Explorer missions, which made measurements like those of CSSWE, the latter has provided critical new information on the dynamics and transport of relativistic electrons and protons in the radiation belts. That, in turn, is helping to improve our models for predicting space weather threats for both robotic space missions and human exploration.

After launch, CSSWE underwent an initial 22-day commissioning phase and then collected 155 days of science data (Gerhardt et al., 2014). The CubeSat was thought to be inoperative when contact could not be reestablished after 7 March 2013. However, after 103 days of communication blackout, the CubeSat came back to life in a designed "Phoenix Mode" on 18 June 2013. Despite the hiatus, the CubeSat was healthy enough to return to science mode, which it did on 27 June 2013. Data collection then continued until 22 December 2014 when the capacity of CSSWE's batteries had degraded extensively and CSSWE could no longer be powered by them. Although the flight mission came to a "second" end, data analysis and modeling continue using a dataset that consists of 3.5 million points covering approximately two total years.

Of the missions reviewed here, CSSWE touts the most impressive publication numbers. In a series of publications, by not only the students who enjoyed early access to CSSWE data but also now broadly by community members, our understanding of the loss and lifetimes of relativistic electrons in Earth's magnetosphere has grown considerably clearer (e.g., Li et al., 2013b; Blum et al., 2013; Zhang et al., 2017). Alone and even more substantially in combination with other missions (Schiller et al., 2014, 2017; Jaynes et al., 2014; Baker et al., 2014, 2021; Li et al, 2015, 2017a; Xiang et al., 2016; Cliverd et al., 2017), the legacy CSSWE data set continues to prove its scientific merits. Owing to its design and flexibility, it not only achieved the mission's full science objectives during the main mission but continues to yield fruit after the mission's completion. For example, a *Nature* paper by Li et al. (2017b) discovered and quantified new phenomena using CSSWE observations, leading to a deeper understanding of the physics of Earth's inner radiation belt, in particular the process by which cosmic rays contribute to the electron radiation belt. That discovery paper led to another by Zhang et al. (2019) who explored

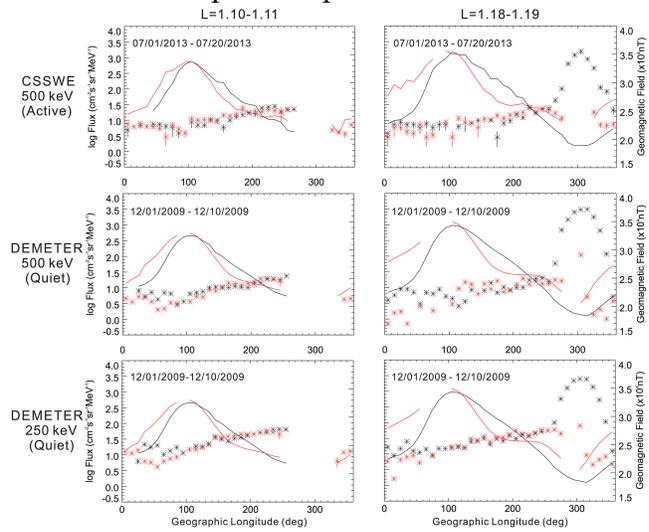

*Figure 3. Electron fluxes (asterisks) as a function of geographic longitude at L = 1.10–1.11 and L = 1.18–1.19 from CSSWE and DEMETER measurements (note that the x axis range is different from Figure 1). Data are binned into 10° longitude bins and averaged over an active period for CSSWE and a quiet period for DEMETER. Solid lines are model geomagnetic field strength at satellite location. Black color stands for satellite locations in the Southern Hemisphere (in terms of geographic latitude) and red color in the north. Statistical error bars are in units of flux per square root of N (N is the number of data points of each asterisk) and are visible when N is small. (From Zhang et al., 2019)*

other aspects of this phenomenon. Figure 3, a reproduction of Figure 2 of Zhang et al. (2019), illustrates through the similarity of CSSWE electron measurements with those taken by DEMETER years apart and under different conditions that cosmic ray neutron decay is a likely source of the quasi-trapped electrons in near-Earth space.

CSSWE Lessons Learned

CSSWE was a very successful university-led CubeSat mission. Its data are still being analyzed and modeled, and so it would be wise to pay attention to lessons learned. They are at least three-fold. We summarize those next.

1. Continuity in documentation: This first lesson deals with assuring continuity of documentation. CSSWE was a typical student mission with inevitable high turn-over; ~40% of student team members left the project or graduated after each semester and a similar number of new students joined the team each semester. CSSWE attributes its success, in large part, to document continuity so that new students could quickly catch up to what had been done and to learn from previous students. Because the mission development was run as part of an academic program, the CSSWE student Project Manager and System Engineer checked on document completion; students could only receive class credits upon approval of their respective mission documentation.

2. Launch serendipity: The CSSWE launch opportunity came at essentially the ideal moment. That was as much the result of serendipity as it was by design. There are many other university-based CubeSat missions for which the launch was severely non-optimal, sometimes years away from the most desirable time. Students who were intimately involved in the final preparations for the mission might have been long graduated. In that case, it is difficult for new students long disconnected from hands-on design experience to be successful when the mission finally launches under their watch. While one cannot plan serendipitous good fortune, the best one can do is try to manage it. Document continuity is one way to manage, amongst others, launch uncertainty.

3. Robust design: For NSF-funded CubeS~at missions, which are extremely cost-constrained (capped at $300K/year for 3-4 years), most teams including the CSSWE team use Commercial Off The Shelf (COTS), rather than space-grade parts. Such parts are more susceptible to deleterious environmental effects than those on larger spacecraft, which have greater resources and commensurately lower mission risk tolerance. Also, like other CubeSat missions, the CSSWE team had no access to the spacecraft once it was delivered, containerized in its Poly-Picosatellite Orbital Deployer (P-POD), and then stored for nine months before launch. A robust CSSWE design anticipated these factors. The team designed CSSWE to revive after launch even if it launched with a completely dead battery, which it did; the system charged up by solar power automatically. CubeSats with COTS parts are more likely to experience abnormal issues in space than other spacecraft, with vulnerabilities to single-event upsets and latch-up. Indeed, CSSWE experienced numerous abnormalities of these sorts; the team's robust design philosophy compensated for the lack of robust parts and as a result the spacecraft recovered from these events by rebooting itself many times, including during the aforementioned "Phoenix" episode.

## 3.4 Focused Investigations of Relativistic Electron Burst Intensity, Range, and Dynamics (FIREBIRD-II)

FIREBIRD-II Overview

The Focused Investigations of Relativistic Electron Burst Intensity, Range, and Dynamics (FIREBIRD-II) mission (Spence et al., 2012) is one of the early NSF-funded, dual-1.5U CubeSat mission and continues to operate. Professors Harlan Spence of the University of New Hampshire (UNH) and David Klumpar of Montana State University (MSU) serve as the original FIREBIRD co-PIs; Professor John Sample of MSU became a faculty mission leader post-launch. The FIREBIRD missions benefit from mission partners at The Aerospace Corporation and at Los Alamos National Laboratory. FIREBIRD used a blended model relying on two universities and their academic programs and students to design, build, test, and operate the mission, and with government partners and their senior members providing design advice and electronic parts in the form of spares from other programs. The main science payload (FIRE) was built at UNH and involved graduate students; the spacecraft bus (BIRD) was built at MSU and involved graduate students and a large cohort of undergraduate students (Klumpar et al., 2015).

FIREBIRD-II was launched in January 2015 out of Vandenberg Air Force Base as a secondary payload on the NASA SMAP mission. To date, FIREBIRD-II is the longest continuously operating NSF CubeSat mission (perhaps the longest operating CubeSat mission of any type). The FIREBIRD-II mission far exceeded its mission duration goal of several months. Both spacecraft operated fully until November 2019, when a battery issue on one flight unit prevented further science collection. At the time of writing of this paper, that unit continues to operate in an engineering mode while the other flight unit continues to return excellent science data even after seven years of essentially flawless operation (Johnson et al., 2020).

Like CSSWE, FIREBIRD-II also explores the physics of Earth's radiation belts, but in a complementary way. Relativistic electron microbursts appear as short (<100 ms) bursts of intense electron precipitation from the radiation belts measured by particle detectors on low-altitude spacecraft when their orbits cross magnetic field lines which thread the outer radiation belt. While microbursts are thought to be a significant loss mechanism for relativistic electrons, they remain poorly understood, thus rendering space weather models of Earth's radiation belts incomplete. Microbursts are generated when distant conditions in the magnetosphere cause electrons to change their trajectories such that they collide with the atmosphere and are lost, rather than electromagnetically mirroring in Earth's magnetic field and remaining trapped in the belts. This sporadic, short time-scale electron dumping from the radiation belts into the upper atmosphere was discovered decades ago. Beginning in 1992, low-altitude observations from SAMPEX provided insight into the morphology of these electron microbursts. They occur in clusters consisting of many individual microbursts. Single satellites, like SAMPEX or even CSSWE, are unable to discern the spatio-temporal behavior of electron microbursts both at the cluster level and at the individual microburst scale. The two-satellite FIREBIRD mission flying in tandem resolves the spatio-temporal variations of individual microbursts for the first time.

The FIREBIRD mission science goals are threefold, all centered on the physical process of relativistic electron microbursts:

1. What is the spatial scale size of an individual microburst?

2. What is the energy dependence of an individual microburst?

3. How much total electron loss from radiation belts do microbursts produce globally?

FIREBIRD-II Science Achievements

Like CSSWE, FIREBIRD-II has enjoyed both significant longevity and a growing community of users who are using these data in their studies. To date, the FIREBIRD-II mission has met all mission goals. The two FIREBIRD spacecraft flew within tens to ~100 km of one another for several months, allowing sampling across many critical spatial scales (Crew et al., 2016). Figure 4, reproduced from Figure 3 of Crew et al. (2016), illustrates how the two FIREBIRD-II flight units (FU-4, upper panel; FU-3, lower panel) observed the evolving precipitation patterns of electrons while only ~11 km apart along their essentially co-orbiting trajectories. These measurements made early in the mission revealed for the first time both the steady and unsteady nature and scale sizes of the electron precipitation, all vital clues to their origins.

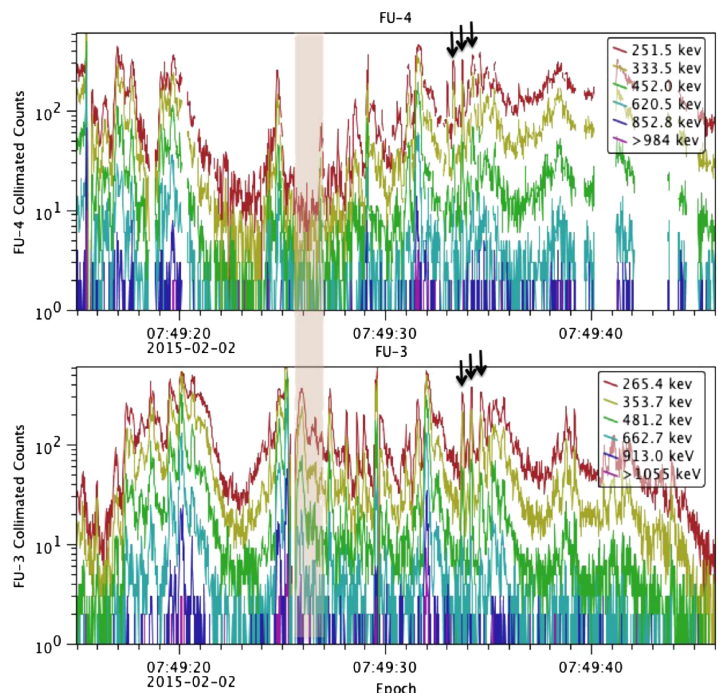

*Figure 4. Microbursts observed during a dawnside pass by both spacecraft. Arrows indicate a clustered region of three microbursts that share the same pattern and timing on both spacecraft, while there are also intervals (such as the shaded one), where the two spacecraft are not seeing correlated microbursts. (From Crew et al., 2016)*

As the two spacecraft drifted further apart over the mission lifetime, the widening separation helped resolve spatial/temporal ambiguity and determined the size of the microburst regions (Anderson et al., 2017; Shumko et al., 2018; Capannolo et al., 2021). This information provides constraints on the physical scattering process and on total radiation belt loss due to microbursts. Measuring electron microbursts with high energy and time resolution (Johnson et al., 2021), FIREBIRD has helped us determine what plasma conditions contribute to the processes that scatter electrons from the magnetosphere into the Earth's upper atmosphere.

While the first two FIREBIRD science objectives constrain the physical processes that generates relativistic electron microburst precipitation, the final objective quantifies the geoeffectiveness and overall space weather impact. That final answer requires cross-track separations of multiple hours of magnetic local time (MLT) ideally on the dawn side, which was not possible within the resources available for the FIREBIRD mission alone. However, FIREBIRD is answering this final highest-level objective and other science questions in combination with other contemporary space and ground assets such as the BARREL balloon mission (Millan et al., 2013), the NASA Van Allen Probes mission, the Japanese Arase mission, ground radar and imaging facilities, and CSSWE.  Examples include studies linking FIREBIRD-II observed precipitation to waves in the source region (Breneman et al., 2017; Capannolo et al., 2019a, 2019b; Chen et al., 2020; Colpitts et al., 2020) and to the pulsating aurora, a ground-based diagnostic of electron precipitation (Kawamura et al., 2021).  We note that plans were made to compare FIREBIRD-II observations with another potential contemporary set of measurements made by a similar instrument (Kanekal et al., 2019), however the NASA CeReS CubeSat mission ended prematurely only five days after launch in late Decemeber 2018.  Finally, in addition, FIREBIRD observations along with related measurements have been used to quantify the effects of electron precipitation on chemistry of the middle atmosphere (Seppälä et al., 2018; Duderstadt et al., 2021), another important consequence of space weather to the neighboring field of atmospheric science. There is broad agreement that low-resource CubeSat missions at low altitude such as FIREBIRD-II and CSSWE and others not described in this paper, such as AC6 (Blake and O'Brien, 2016), have advanced the science associated with energetic charged particles in Earth's magnetosphere (e.g., Fennell et al., 2016).

FIREBIRD-II Lessons Learned

FIREBIRD-II shares many lessons learned from those described earlier.  Like RAX-2, FIREBIRD-II benefited from a reflight opportunity and that is probably the greatest lesson.  The original FIREBIRD-I mission had only partial success; while both of the FIREBIRD spacecraft operated, they did not operate together at the same time owing to a design flaw.  At a small fraction (~20%) of the original mission cost, the same team modified and improved the design and FIREBIRD-II launched a few years later.  That reflight has proven to be wildly successful and it is in no small part the result of having the chance to learn and improve designs with ostensibly the same team.  We note that a third generation of the FIREBIRD-II instrument is slated to fly as part of the NASA-funded AEPEX mission (Marshall et al., 2020), continuing the "fly-learn-modify-refly" cycle.

Because the second launch occurred after a rather long (compared to a typical time an undergraduate spends working on the project) delay after the first, the importance of documentation was also critical.   In the case of FIREBIRD-II (and also CSSWE), many of the student leaders that worked on the development have remained involved as their professional career has evolved; that continuity is also an important component for success.  For those interested in more details of the design lessons learned on FIREBIRD-II, please refer to the published paper on this very topic by Klumpar et al. (2014).

Finally, FIREBIRD is a prime example of a mission whose instrumentation generates far more data volume than could be telemetered to the ground within mission resources. Because the mission science required the identification of comparatively rare features in the data, the mission team developed two approaches for finding the proverbial needles in the haystack.  First, they developed on-board algorithms that attempted to identify microbursts automatically in a well-defined manner.  Though tested on the ground with other data, this algorithm proved to be unsuccessful in flight when using raw, unprocessed FIREBIRD data. A second approach ultimately employed the scientist-in-the-loop mode to identify which data intervals to download based on a grossly time-averaged data product.  While this approach worked, the process is inherently labor intensive and imperfect in always identifying the best intervals.  Given this significant impact to science return, an important lesson learned from the experience is that future such missions urgently need strategies for more data return, both through more robust onboard processing and data down selection and through improved communications approaches.

### *3.5 Miniature X-ray Solar Spectrometer (MinXSS)*

MinXSS Overview

The Miniature X-ray Solar Spectrometer (MinXSS) mission (Mason et al., 2016; Woods et al., 2017) was NASA's first-launched science-oriented CubeSat and another recent example of a highly successful application of a smallsat platform for space weather-related research, complementary to the NSF missions described above. Unlike the previous four missions, MinXSS was a remote sensing solar mission rather than an *in situ* geospace mission, which necessitated significantly different hardware considerations (for example, constant solar pointing helps with power, but complicates thermal issues and requires fine attitude control).   MinXSS-1 was deployed from the International Space Station in May 2016 and operated successfully for nearly a year, de-orbiting in May 2017. Its primary objective was to measure the solar spectral irradiance in soft X-rays (SXRs; ~0.5–30 keV, or ~0.04–2.5 nm) to determine the wavelength-dependent energy flux incident on Earth's ionosphere, thermosphere, and mesosphere (ITM). Solar SXRs are the dominant drivers of dynamics in the D- and E- regions of the ionosphere (Sojka et al., 2013; 2014), as well as of various NOx-related photochemical reactions within the ITM (Baily et al., 2002); the specific dynamics are strongly dependent upon the spectral distribution (amount of energy at a given wavelength), particularly within the 1–5 nm band that is highly variable with solar activity (Rodgers et al., 2006). Thus, measuring the SXR spectral energy distribution with sufficient resolution to constrain the inputs to these energetic processes is critical to understanding solar forcing of ITM dynamics.

MinXSS Science Achievements

MinXSS-1 measured the SXR spectral irradiance shown in Figure 5 with a resolution of ~0.15 keV FWHM (quasi-constant in energy, variable in wavelength as $\Delta\lambda=hc\Delta E/E^2$) and cadence of 10 s (taken from Figure 15 of Moore et al. 2018) using a commercial off-the-shelf (COTS) miniaturized silicon drift detector (SDD), achieving the best spectral resolution to date over this broad passband, and over dynamically relevant timescales. For example, MinXSS measurements reveal the wavelength-dependent energy distribution of the broadband (0.1–0.8 nm) integrated irradiance observed by GOES X-ray Sensor (XRS) photometer, and first results have suggested that the GOES-reported irradiance levels may be inaccurate at low flux levels (Woods et al., 2017). This has a direct significance for solar plasma temperatures inferred from these measurements (e.g., Caspi et al. 2015), which are often used to estimate the SXR flux incident on the ITM to drive atmospheric models. These results are being further explored through additional on-going analyses (e.g., Reep et al., 2020).

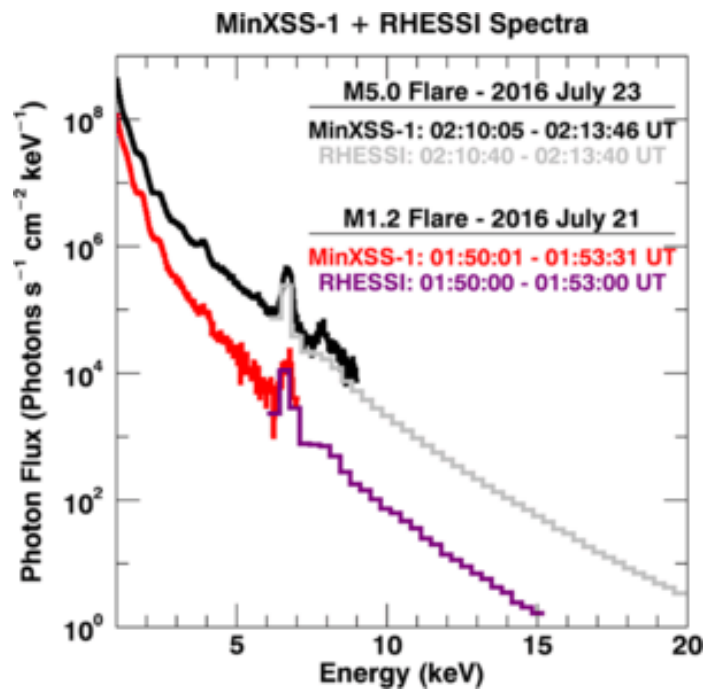

The MinXSS solar spectra overlap the lowest energy part of the RHESSI spectra as shown in Figure 5. The unique MinXSS spectral range of 1-6 keV has several emission lines from Mg, Si, S, and Fe, which are providing new information about the elemental abundance changes for studying flare energetics and nanoflare heating in solar active regions. We note that technologies and tools developed for MinXSS have dual use with other NASA Science Mission Directorate X-ray missions, including solar observations made by the Astrophysics Division NuSTAR mission (Hannah et al., 2016; Grefenstette et al., 2016) and by Solar Dynamics Observatory (Aschwanden et al., 2017). Furthermore, MinXSS data have been incorporated into the second version of the Flare Irradiance Spectral Model (Chamberlin et al., 2020), which has been used in many related studies of solar flares (Chamberlin et al., 2018), and, applicable to differential emission measure for MinXSS and other X-ray observations (McTiernan et al., 2019; Plowman and Caspi, 2020).

A follow-on mission, MinXSS-2 (Mason et al., 2020), launched into sun-synchronous polar orbit in December 2018 on the Spaceflight Industries SSO-A "SmallSat Express" launch, to continue making these important measurements. MinXSS-2 included an upgraded detector and had an estimated 4-year mission lifetime but suffered an electronics failure in late January 2019; a hard reboot was commanded from the ground to clear an issue with the onboard SD-card, but communication with the spacecraft ceased immediately thereafter. A variant of the MinXSS-2 detector with a nested aperture design (Schwab et al., 2020), optimized for moderate solar activity anticipated during the rise of the current solar cycle, will launch as a hosted payload on the INSPIRESat-1 CubeSat, anticipated for Q1 2022 (Chandran et al., 2021). The focus for MinXSS-3 instrument science is the study of solar active region evolution and continuation of the studies about flare energetics and impacts on Earth's ionosphere.

MinXSS Lessons Learned

Besides its miniature detector, one of the primary enabling technologies for MinXSS-1's groundbreaking success was the XACT attitude determination and control system (ADCS) from Blue Canyon Technologies (BCT), that provided high-precision, stable pointing in a compact, ½U form factor (Mason et al. 2017). The XACT was critical to MinXSS success because of the need for fine-Sun pointing. MinXSS-1 was the first flight of the XACT, demonstrating better than 10 arc-sec pointing capability, and the XACT has subsequently been flown on many missions.

The rest of the MinXSS hardware, including its detectors, was largely built from COTS components (albeit with student-led design and assembly). One of the key lessons learned from MinXSS was that, despite CubeSats' small size, low cost, and use of COTS components, a rigorous testing plan is essential to reduce risk and increase the chances of success. MinXSS followed the same general testing strategy as larger missions, including design reviews, environmental (vibration and thermal vacuum) testing, and end-to-end testing prior to shipment. Although the tests were scaled down to appropriate levels of effort and risk tolerance for CubeSats, they were crucial in finding and fixing potentially fatal flaws before launch, and, validating that MinXSS's design and performance were robust and reliable for a year-long mission in space. The importance of this lesson is especially evident considering the fate of MinXSS-2. The electronics failure for MinXSS-2 likely occurred in the SD-card used to store mission data; the operational software was encoded in non-volatile firmware and was unaffected. We note that MinXSS-2 mission had higher orbital inclination and higher altitude than MinXSS-1, so the radiation environment for MinXSS-2 was much harsher. MinXSS-2 included a hard-reboot circuit that was added to help recover quickly from single-event upsets and/or latch-ups like these, a lesson learned from the MinXSS-1 mission. However, design limitations in the hard-reboot circuit and certain software interactions resulting from the SD card failure likely led to a watchdog timeout condition on system startup as the system attempted to initialize the SD-card interface, causing the spacecraft to become unresponsive. This failure condition was able to be reproduced in the ground-based MinXSS-3 flatsat setup with a corrupted SD-card installed. In hindsight, a watchdog timer duration of more than 60 seconds allows recovery for this SD-card configuration. This MinXSS-2 failure highlights the need to test failure modes, particularly for large and/or highly susceptible parts, and interactions between watchdog timers and reset circuits. Budgetary constraints are the biggest hurdle to such testing, so prioritization and optimization of testing is key. Of course, the rapid-and-inexpensive-turnaround nature of CubeSat reflights mitigates, to some extent, this requirement, as lessons learned can be implemented on a new build and flown again at relatively low cost, as evidenced by both RAX-2 and FIREBIRD-II and on the next generation MinXSS-3 sensor planned for launch in early 2022.

One of the key challenges for MinXSS science and operations was its use of "ham" UHF radio frequency communications. This was chosen for its flight heritage (from CSSWE), low cost, and relative simplicity (including in frequency licensing from the FCC). However, this limited downlink rates to only 9600 baud – less than 1 KB/s after encoding overhead. With only one ground station, total downlink capacity was theoretically only ~1 MB/day, and other operational considerations (required command uplinks, radio interference, etc.) limited actual average capacity to significantly less. In contrast, raw data generation exceeded ~30 MB/day (including housekeeping but excluding diagnostic data). Even with custom on-board compression of the science data, total generation was a few MB/day, and only ≲5% of the total science and housekeeping data was able to be downlinked over the 1-year mission. Despite the groundbreaking nature of MinXSS observations, this data rate limitation imposed by the UHF implementation significantly restricted available science and introduced additional operational complexity to prioritize downlink of critical observations (e.g., of solar flares) on top of the added flight software complexity for on-board compression. The follow-on MinXSS-2 mission used an additional ground station in Alaska and thus benefited from higher ground station visibility from its sun-synchronous polar orbit but had the same fundamental limitations and thus was also restricted to only ~10–20% data recovery. MinXSS-2 radio and ground stations were configured to switch to 19200 baud to increase the data recovery to ~40%, but that goal was not achieved due to its SD-card anomaly impacting the MinXSS-2 operations.

Higher data rates are now routinely achievable from smallsat platforms, with S-band transceivers and X-band transmitters with flight heritage already on the market, and X- and Ka-band transceivers in development. Smallsat-compatible optical (laser) communications terminals are also being developed. Although more expensive than "ham-radio" UHF options, these solutions would allow complete data capture from a MinXSS-like mission while allowing complexity reduction by obviating the need for on-board compression and downlink prioritization schemes. The complexity reduction may be at least partially offset, though, if the more capable transmitters require increased power to operate. Such a power impact could increase the complexity of both the power system and the overall thermal design. We note that MinXSS benefited from a rather simple thermal design (Mason et al., 2018). Other factors may offset complexity reduction, such as frequency licensing. For missions with larger data generation (e.g., CubIXSS and the other mission concepts presented in the Caspi et al., 2021 companion paper), these higher data rate solutions are imperative to enable breakthrough science to enhance space weather research and operations; a thorough trade study on this topic would benefit the entire CubeSat community.

**4 Common Themes**

Several common themes emerge when assessing the scientific success of these missions. First and foremost, missions even as small as a single CubeSat can contribute significantly to the space weather enterprise. In all instances, these small missions remained focused on one aspect of space weather. Consequently, they could address comparatively more narrowly focused science goals than missions that are much larger in scope. Those science goals are often no less worthy than multiple goals sought by larger mission. Indeed, one could argue that a more focused science goal better sharpens and limits mission need, resulting in a more cost-effective approach to answering isolated problems. Even with such limits, many of the missions reviewed have not only answered the narrow goals they were designed for, but, owing to their new focused capability, have also revealed new science questions that motivated future missions or made new discoveries.

Another common theme is the limits on science return imposed by resource-limited communications. On the missions described, it is not uncommon to retrieve only a few percent of the data collected by onboard instrumentation. This is in stark contrast to physically larger missions that generate sufficient power to operate highly capable radio systems and transmit typically a larger fraction of collected data. As new technologies and approaches emerge (such as shared, standardized ground stations) in the small satellite community, the communications return gap between small and large missions continues to narrow. There may even be opportunities to more deliberately leverage, increase, and/or support the amateur "ham" community. We note that NASA's deep space missions are also subject to this same issue of comparatively low data rates. CubeSats are inherently more risk-tolerant compared to billion dollar class planetary missions, for example, and are thus an excellent platform to explore novel solutions to bandwidth limitations common to many mission classes.

In virtually all cases, another unifying theme is that these small missions allow investigators to learn from mission imperfections or flat-out mistakes. The initial missions themselves are low cost (typically ~$1M for NSF missions); the cost to rebuild and refly the same mission with modest redesigns is typically a small increment of that initial investment (~$200k for NSF missions). The ability for students to learn from a design flaw, to modify and correct it, and to fly it again successfully in a short time is as invaluable to their learning as it is incremental in cost. While this approach is impractical for large, complex, costly missions, it is proving to be an effective opportunity for these very small missions. The risk of mission failure is mitigated by the opportunity to inexpensively fly, learn, modify, and refly. We note that while NSF missions are ~$1M, NASA CubeSats are now routinely ~$5M and growing (for example, CubIXSS is ~$7M, and another recently selected mission, PADRE, is ~$9M). This reflects two realities: that increasing the probability of achieving high-quality science does require additional investment; and that NASA recognizes the importance of CubeSats in filling various observational and technology-development gaps. The higher costs are not merely due to Phase E being more robust. For instance, owing to NASA's different focus, these missions tend to employ more professionals than students, and often use more expensive COTS components (e.g., space-rated COTS) than the NSF program. Even at this higher cost, NASA CubeSats are still 1-2 orders of magnitude cheaper than NASA Explorers, and still benefit from lower re-flight costs.

The need to document work so underpins the entire fly-learn-modify-refly cycle that it deserves mention as a common theme. Agile engineering works because there exists a formal process of which documentation is a critical element. Documentation is especially important in projects that involve evolving teams of students often disconnected and non-overlapping in time. The next team benefits not only from understanding what development came before their work, but also what was learned and communicated forward through documentation. All the missions described above were executed at institutions who benefited from their own internal engineering processes, most commonly developed over time through prior larger NASA programs. However, given the relatively smaller budgets associated with CubeSats, such fuller engineering processes (including documentation) had to be tailored so that acceptable mission risk is balanced against available funding. Regardless of mission scope, documentation remains one of the most important tools to foster continuous improvement in any engineering process.

Finally, another common outcome for these small missions is a high scientific return on investment. This outcome is underscored by quantifying science productivity (as measured by number of peer-reviewed publications) normalized by mission cost (which are all ~$1M). Table 1 summarizes that metric for the five missions described in Section 3. To construct Table 1, we

counted every peer-reviewed paper with a Digital Object Identifier (DOI) published in scientific journals, conference proceedings, and book chapters; we did not count Masters and PhD theses or any publications lacking a definitive DOI. These missions on average produce 2.0 peer-reviewed publications per year per million dollars. Of course, science productivity is not the only goal of these NSF awards. Student training is key as well and that is not accounted for in this metric; because most CubeSats to date have been largely implemented with students at universities, there is also an expectation that there will be more PhDs produced (and master's degrees) per $M than from traditional large satellite projects.

Table 1 also compares the CubeSat/smallsat mission levels of productivity to those of larger NASA missions with other metrics. We use publicly available (mission web sites, NASA Senior Reviews, etc.) values for numbers of publications for representative Heliophysics missions including a Small Explorer (SAMPEX), a Medium Explorer (THEMIS), a strategic mission (Van Allen Probes), and a flagship mission (Magnetospheric Multiscale). Clearly, given the broader scientific scope of the larger missions, their larger science teams, and the typically more substantial investments in the science payload and mission science phases of the large missions, the total of CubeSat publications per mission on average is a small fraction (~1.5%) of that produced on average by the large mission; that is still a very small fraction (~2.5%) even when normalized by years since mission launch.

*Table 1. Comparison of publication productivity metrics of missions by scale*

| Mission Category | Funding Agency | Mission Name | Prime Mission Cost (M$ FY22)* | Years Since Launch (YSL) | Peer-reviewed Publications | Weighted Publication Impact Factor | Peer-reviewed Publications per YSL | Peer-reviewed Publications per YSL per M$ |
|---|---|---|---|---|---|---|---|---|
| CubeSat | NSF | **DICE** | 1.3 | 10.3 | 9 | 1.46 | 0.9 | 0.7 |
| CubeSat | NSF | **RAX-2** | 1.3 | 10.3 | 12 | 2.99 | 1.2 | 0.9 |
| CubeSat | NSF | **CSSWE** | 1.3 | 9.3 | 25 | 6.41 | 2.7 | 2.0 |
| CubeSat | NSF | **FIREBIRD-II** | 1.2 | 7.0 | 19 | 3.52 | 2.7 | 2.2 |
| CubeSat | NSF / NASA | **MinXSS** | 1.2 | 5.7 | 17 | 4.54 | 3.0 | 2.4 |
| | | *Average* | | 8.5 | 16.4 | 3.8 | 2.1 | 1.6 |
| | | | | | | | | |
| SMEX | NASA | **SAMPEX** | 72 | 29.5 | 2000 | - | 67.8 | 0.9 |
| SMEX | NASA | **IRIS** | 99 | 8.5 | 442 | | 51.9 | 0.5 |
| MIDEX | NASA | **THEMIS / Artemis** | 230 | 14.9 | 1699 | 3.54 | 113.9 | 0.5 |
| Strategic | NASA | **VAP** | 670 | 9.4 | 893 | 4.02 | 94.8 | 0.1 |
| Flagship | NASA | **MMS** | 1474 | 6.8 | 681 | 4.43 | 99.6 | 0.1 |
| | | *Average* | | 13.8 | 1143.0 | 4.0 | 85.6 | 0.4 |

* Using NASA inflation tables from cost at year of launch to FY22

However, Table 1 demonstrates that the smallsat missions outperform these larger missions with another publication metric (last column), namely the total number of publications normalized by total mission cost (and time since launch to account for different length science phases). By this measure, the CubeSat missions on average produce almost four times the number of publications per dollar per year. There are many reasons for this, not the least of which is launch costs, the larger costs required to assure extremely low risk, and more complex spacecraft and missions needed to achieve more challenging science goal for larger missions. However, what the smaller missions lack in terms of net publication production (columns six and eight), they make up in terms of cost value per publication (column nine).

We note an extremely important caveat regarding mission costs listed in Table 1. All missions ultimately benefit from funded development that precedes that mission. Instrument development programs provide funding to move an instrument concept from a low technical readiness level (TRL) to one that has demonstrated enough design maturity and level of risk to be selected. Subsequently, many of these instruments' TRLs increase even higher through sub-orbital programs. None of these development costs are included in Table 1 nor would it be easy to do so. Such a full cost accounting of missions would be very challenging, recognizing that all missions leverage prior efforts to some degree. Despite those caveats, we adopt the accepted mission costs as a means for simple comparison. Each mission leverages prior development costs in different ways and at different levels. In the case of CubeSats, their low mission cost almost assures that their cost is a far greater underestimate of the true mission costs compared to the larger missions. Indeed, as noted above, we can point to how the CubeSat community leverages other support outside the noted mission costs, including: the unfunded benefit of the amateur "ham" radio community; the unfunded benefit of national experts and institutional capability; and the uncosted benefit of national programs that provide launch opportunities along with larger missions. This is not meant to be a criticism, but rather is an expression of how programs interrelate in this "ecosystem".

A common impression is that CubeSat mission publications tend to generate less scientific impact compared to those from larger missions. As was noted in the National Academy Report, CubeSats have a higher proportion of technical papers compared to science publications, in part because CubeSat engineering is still evolving; that is a feature of having overall fewer publications with a similar number of required technical papers compared to large missions. We test that quantitatively with a different metric shown in column seven (note this metric is not calculated for the SAMPEX and IRIS missions as we did not have a definitive list of mission publications).

For each of the publications appearing in column six for each mission, we identified the publication journal. We then tallied the number of publications appearing in all journals for each mission. To gauge the net impact of a mission's publications, we produced a weighted publication impact factor. We computed a weighted sum of the products of the numbers of publications in any given journal times that journal's impact factor. We used values of journal impact factors provided in the 2021 Journal Citation Report (JCR) (https://impactfactorforjournal.com/jcr-2021/) published by Clarivate Analytics, a Web of Science group. JCR defines the 2021 impact factor as the sum of all citations from 2019 and 2020 divided by the total number of papers published in that journal in 2019 and 2020. For publications that appeared in journals for which no impact factor was available (for example, some conference proceedings or book chapters), a low but non-zero value of 0.1 was used. Each mission weighted sum was then divided by the total number of publications for each mission to yield an overall weighted publication impact factor.

While this simple approach has shortcomings (e.g., it does not account for how a journal's impact factor changes with time, it assumes that all publications in a given journal have the same scientific impact, it does not account for actual citations of the papers nor the positive or negative character of those citations, etc.), it does provide at least a reasonable quantitative measure of the relative quality of journals, as assessed by an independent group, in which the eight space science missions published their results. Likely contrary to conventional wisdom, this metric quantifies that CubeSat missions hold their own in terms of weighted publication impact factor relative to the larger missions, with comparable on average (impact factors of ~4.0) and even higher values in individual cases compared to the larger missions. For reference, JCR reports 2021 impact factors of ~2.8 for the *Journal of Geophysical Research – Space Physics* and ~4.7 for *Geophysical Research Letters* (the two most frequent journal publications in our survey); apropos for this topic, JCR reports an impact factor of ~4.5 for the *Space Weather Journal*. If nothing else, Table 1 demonstrates that it can be misleading to compare mission success by only comparing total number of publications per mission. The value and outcomes of missions require deeper inspection.

We do not wish to imply from Table 1 that CubeSats should or shall ever *replace* large missions – their strength lies in exploring highly targeted science questions, particularly those requiring multi-point measurements from constellations (Caspi et al. 2021, Verkhoglyadova et al. 2020), and in exploring discovery space as pathfinders to larger missions. Instead, CubeSats should be considered as highly complementary, filling gaps that may be infeasible through larger missions or pioneering research avenues that reduce future risk on larger missions. As such, they should be nurtured as part of a robust research satellite "ecosystem." There are many times when mission scope and mission implementation demand large spacecraft, significant investment, and thus low risk. Table 1 does suggest however that a balanced ecosystem need not be one in which smaller missions dilute the scientific impact of an overall mission portfolio but rather is one in which there is considerable value and rationale for implementing missions over a broad range of sizes and scopes.

Even though scientific productivity is already measurably high for these small missions, we believe that the return could be even higher, particularly for the NSF-funded missions. The NSF funding model supports a team to design, develop, integrate, test, deliver, and operate a CubeSat on a mission that has space weather science goals. With a strict cost-cap of $1.2M (~$300k/year for up to 4 years), typically, little funding remains after those activities to conduct scientific research, particularly any beyond a typically very short prime mission phase. For missions that operate successfully, science productivity depends on a PI writing and being awarded a new grant to conduct science; this process has a built-in delay, at best, and an unsure outcome. In the worst case, a team may not have their data analysis proposal funded (even if fundable). In some instances, science outcomes rely on other related awards or internal funds to advance the cause. In the future, cost savings through increased standardization of commercial bus / bus systems could be invested toward payload development and science analysis to foster even greater CubeSat mission science return within the same cost cap.

The NASA mission model provides an alternative approach. The NASA CubeSat program (now under Heliophysics Flight Opportunities for Research and Technology, H-FORT) has a larger available budget for new starts, with no explicit cost cap, so PIs can include funding specifically for research and data analysis. Indeed, the NASA H-FORT program explicitly states that "budgets are expected to cover complete investigations" and requires "data analysis, data archiving, and publication of results". Recent missions have been funded in the $7M+ range, with durations up to 5 years. The larger budget enables significant analysis even during the operational period, and NASA also has specific mechanisms in place that can enable funding extensions for missions that are still operational and yielding good science return at the end of their nominal proposed period of performance. In the future, the NSF might consider funding CubeSat missions

with a more deliberate attention to science funding, perhaps as an option in the original proposal, should missions demonstrate sufficient early success.

**5 Conclusions**

Current scientific investigations span across solar, space physics, space weather, and atmospheric research but evidence from the successful projects so far strongly suggests that the future of scientific CubeSat projects is only limited by imagination. Additional measurements from space are crucial not only to address many unsolved science problems but also to solve critical societal problems, such as climate change; land use and resource management; pollution and disaster monitoring; communication; and space weather. CubeSat missions can help provide these and, in particular, offer a realistic and low-cost means of realizing widespread use of constellations of many satellites to address global system science, which remains a potential game-changing goal for many science applications, not least including space weather.

**Acknowledgements**: H.S. was partially supported by NSF grant (ATM-1035642) and also by RBSP-ECT UNH funding provided by JHU/APL Contract No. 967399 under NASA's Prime Contract No. NAS5-01072. A.C. was partially supported by NASA grants NNX14AH54G, NNX15AQ68G, NNX17AI71G, and 80NSSC19K0287. T.N.W was partially supported by NASA grants NNX14AN84G and NNX17AI71G. RAX was funded by the U.S. National Science Foundation, Grant ATM-0838054. Additional funding for this work was provided by the U.S. Department of Defense through a National Science and Engineering Graduate (NDSEG) Fellowship. The authors also wish to thank the organizers of the 1st International Workshop on Small Satellites for Space Weather Research & Forecasting (SSWRF) for their efforts and for motivating this manuscript and its companion papers, and NSF award 1712718 for workshop funding support. No new data were used or generated in preparing this manuscript; all cited data from prior publications can be found in the respective cited works. Lists of peer-reviewed publications used to compute the Weighted Publication Impact Factors in the lower section of Table 1 can be found at the publicly available mission websites for THEMIS (http://themis.igpp.ucla.edu/publications.shtml); Van Allen Probes (https://rbspgway.jhuapl.edu/biblio?keyword=Van%20Allen%20Probes); and MMS (https://lasp.colorado.edu/galaxy/display/mms/MMS+Publications). The Weighted Publication Impact Factors in the upper section of Table 1 are computed using the publications cited in this paper for each CubeSat mission. Journal impact factors for 2021 used in Table 1 can be found at: https://impactfactorforjournal.com/jcr-2021/. "